\documentclass[10pt,a4paper]{article}
\usepackage{amsmath}
\usepackage{amsthm,amssymb}
\numberwithin{equation}{section}
\usepackage{graphicx}

\title{\Large \bf  Primacy  \& Ranking of UEFA  Soccer Teams \\from  Biasing Organizing Rules   }

\author{ \large \bf  Marcel    Ausloos$^{1,*,2,}$,  Adam Gadomski$^3$,    Nikolay K. Vitanov$^4$  
 \\ \\$^{1,*}$  Royal Netherlands Academy of Arts and Sciences\footnote{Associate researcher}, \\Joan Muyskenweg 25, 1096 CJ Amsterdam, The Netherlands \\  
\\ $^2$ GRAPES, rue de la Belle Jardini\`ere 483, B-4031 Li\`ege, \\Federation Wallonie-Bruxelles, Belgium  \\  email: marcel.ausloos@ulg.ac.be\\
 \\ $^3$ University of Technology and Life Sciences, \\ Department of Physics, Institute of Mathematics and Physics, \\PL-85-796 Bydgoszcz, Poland\\email:  agad@utp.edu.pl; agad@atr.bydgoszcz.pl  \\
\\$^4$  Institute of Mechanics, Bulgarian Academy of Sciences, \\
Acad. G. Bonchev Str., Bl. 4, BG-1113 Sofia, Bulgaria\\email: vitanov@imbm.bas.bg}

\begin{document}
\maketitle
\begin{abstract}
A question is raised on  whether some implied regularity or structure, as found in soccer team ranking  by  the  Union of European Football Associations (UEFA), is due to implicit game result value or score competition conditions.  The analysis is based on considerations about  complex systems, i.e. searching whether power or other simple  law fits are appropriate to describe some internal dynamics. It is observed that  the ranking is specifically organized: a major class made of  a few teams emerges after each game season.   Other classes which  apparently have regular sizes subsequently occur.  
Thus,  the notion of  Sheppard primacy  index is envisaged to describe the findings.  Additional primacy indices  are  discussed for enhancing the features.  These measures can be  used to sort out peer classes in more general terms. A very simplified toy model containing  ingredients of the UEFA ranking rules  \underline{suggests}  that  such peer classes  are  an extrinsic property of the ranking, as obtained in many nonlinear systems under boundary condition constraints.  
\end{abstract}

\section{Introduction}\label{Introduction}
\par
Nonlinearity   and complexity 
\cite{compl1}-\cite{compl4} are common features of a  large number of systems studied in  modern science \cite{syst1}-\cite{kwapien2012physical}. They are often investigated by  nonlinear dynamics  methods  \cite{tsallis2012entropy}-\cite{str}. 
 In the last decade or so, these methods have been applied to many social, economic, and financial systems  \cite{vesp}.
In many complex systems,  researchers have detected the existence of power laws,  for  a large variety of  characteristic quantities.   Such power laws have become very useful tools  for studying complex systems because the functional relations  can  indicate that the system is controlled by a few rules that propagate across a  wide range of scales \cite{stanley}-\cite{g2}.   

For example, ranking analysis has   received much attention  
   since  Zipf \cite{z1}  observed  that a large number 
of  size  distributions, $N_r$, can  be approximated by a simple  {\it 
scaling (power) law} $N_r = N_1/r $,   where  $r$ is the ranking parameter, with 
 $N_{r } \ge N_{r+1}$, (and obviously  $r<r+1$). 
This  idea has led to a flurry of log-log diagrams showing a straight line through the displayed data.  More generally, one considers the  so-called rank-size scaling law:
  \begin{equation}\label{Zipfeq}
y_r = \cfrac{a}{r^\alpha} \;.
\end{equation}
where the scaling exponent  $\alpha$ is considered   indicative of whether  the size distribution  $y_r $ is close  to  some optimum  (= equilibrium) state \cite{z1}, i.e. when $\alpha=1$.  
The amplitude $a$ can be estimated from the normalization condition. For the discrete distribution,  Eq.(\ref{Zipfeq}), $a\simeq r_M/ \zeta(\alpha) \sim r_M/2$, where $r_M$ is the largest value of $r$ and $\zeta(\alpha)$ $\equiv$$ \sum_{k=1}^{\infty} k^{-\alpha} $ is the Riemann zeta function \cite{Titchmarsh}.

This  scaling  hypothesis might be applied  in sport competition ranking, - though the number of scales  would obviously be  finite. Nevertheless, measurements or ranking in sport competitions,  while  frequently reported in the media,  lack the necessary descriptive power  which   studies of complex systems  usually present  and require in physics investigations.   An analysis of  data from a specific nonlinear complex system,   the  Union of European Football Associations (UEFA) team ranking,  is here below presented   as a specific  and interesting modern society example. Deviations
at low and high rank $r$, from the empirical fits to a single power law  indicates the existence of different regimes.
 For completeness, the UEFA rules leading to its  team ranking coefficient are  briefly recalled in Appendix A.  They suggest extrinsic biases.

 Introducing an indirect "primacy measure"   based on Sheppard hierarchy index  \cite{s1},  - (see Appendix B for  a review of  Sheppard's original index), it is  found that  UEFA teams can be organized in several  well defined classes. Whether or not this  is related to or could be used for weighting performances is speculation, but cannot be {\it a priori} disregarded. However, practical applications cannot be recommended  from  this report, because    any recommendations would be  outside the  present scientific aims, i.e. some search for  structural features in a social complex system, - team ranking.

The paper is organized as follows.  In Sect. \ref{stateoftheart}, a brief review of the literature on ranking,  and in particular for  soccer teams,  is presented.  
 The data analysis is performed  in Sect. \ref{UEFAdata}.  Simple empirical laws are briefly reviewed in order to introduce possible fit laws.
The rank-size relationship, Eq.(\ref{Zipfeq}),  is assessed for  UEFA teams. Various fits  point to features, emphasized in figures  displaying various empirical  laws.
In Sect. \ref{primacy},  the  hierarchy inside the top  classes is  further analyzed,  starting from the conventional  primacy index of Sheppard \cite{s1},   through additional 
measures of primacy.  
 Several remarks  serve as conclusions, in Sect. \ref{conclusions}.

A very simplified "toy model" is  numerically discussed in Appendix C,  - in order to show that under UEFA team ranking  rules, gaps necessarily emerge between classes of teams, -  thereby suggesting that  the features mimic thermodynamic dissipative structures in finite size systems.

\section{State of the Art}\label{stateoftheart}

The rank-size relationship, Eq.(\ref{Zipfeq}),   has  been frequently identified and sufficiently discussed in the literature 
to allow us  much of the present investigation to be based  on such a simple empirical  law.  
This may
be "simply" because the rank-size relationship can be reached 
from a wide range of specific situations.
Indeed, Zipf's law, Eq.(\ref{Zipfeq}) with$\alpha=1$, and its generalizations,  can be obtained in different models: one  example is tied to the maximization of 
the entropy  concept \cite{PhA391.12.767ranksizescaling}; another stems from  the law of proportionate 
effect, so called Gibrat's law  \cite{gibrat}.  Recall  that 
  Gibrat`s law describes an $evolution$ process, supposing  that the growth rate of  something   is independent of its  size and  previous rank.   

Since the  specialized  literature on  team  ranking, is not  common in the physics literature,  
  a brief "state of the art"  is  presented here below, only pointing to {\it a few} publications:   
\begin{itemize}
 \item
 Stefani   \cite{JAS24.97.635worldsportsratingStefani} 
pioneering survey of the major world sports rating systems in  1997 

\item  Cassady  {\it et al.}    \cite{interfaces35.06.497NCAA}  discussion of  a customizable quadratic assignment approach for ranking sports teams in 2005, 
  \item  Churilov and Flitman's  \cite{COR33.06.2057} proposal that the Data Envelopment Analysis (DEA) model for producing a   ranking of   teams  or countries, - like   in the  olympics  games,  in 2006,  
  \item    Broadie and  Rendleman's \cite{owgr_20120507_broadie_rendleman} audacious question of  whether   the   official World golf rankings  are biased.
    \end{itemize}
 
 In one  highly  conclusive paper,  entitled  {\it Universal scaling in sports ranking} \cite{NJP14.12093038scoreprize}, 
    the authors  
studied the distributions of scores and prize money in various sports,  showing  that  different sports share  similar trends in  scores and prize money distributions, whence pointing to many implications. In a related  study,  Pilavci's  Masters thesis  (see  \cite{pilavci2011pitch} and references therein)  evaluated economic, demographic, and traditional factors which affect  soccer clubs on-pitch success in UEFA.

Other ranking studies, for  National Collegiate Athletics
Association (NCAA) and  for National Football League (NFL)  teams  
  \cite{JSE10.09.582weightranking,NFLranking}, have been based on subjective  considerations or indirect measures.
 
Specifically for  soccer  $ ranking$ themes,  the following studies should be mentioned:
     \begin{itemize}
 \item  Kern and   Paulusma \cite{DAM108.01.317}  discuss FIFA rules  complexity  for   competition outcomes 
and   $team$ ranking  in 2001;
  \item Macmillan and  Smith  \cite{JSE08.07.202explainingrankingsoccercountries}   explain such a $country$  ranking in 2007; 
  \item  Ausloos et al. \cite{IJMPCFIFAMARCAGNV}   compare  the  \underline{country  FIFA ranking}, - based on games between national  squads,  and  the \underline{country UEFA ranking}, based on  \underline{team}  game results;
  \item  Constantinou and Fenton \cite{pi-ratings}    determine   the level of "ability" of  (five English Premier League seasons) soccer teams by  ratings based on the relative discrepancies in scores between adversaries. 
    \end{itemize}
    
Other papers with some $soccer$  related content  can be mentioned for  completion:  e.g.,  the flash-lag error effect in soccer games \cite{Ranvaudflagerrors},    the 'best' team win frequency \cite{JAS36.09.soccer}, 
  the upset  frequency  as  a  measure of competitiveness  \cite{0608007},
 the "evaluation" of goals scored in Euro 2012 \cite{IJSS3.13.102goalsEuro012},     the  goal distributions  \cite{EPJB67.09.459Janke,EPhL78.07.58002Janke},   
the relationship between the time of the  first goal in the game and the time of the second goal \cite{12076796Nevo}, 
  the goal difference as a better measure for the overall fitness of a team \cite{EPJB67.09.445Heuer},     the general dynamics of soccer tournaments \cite{EPJB75.10.327soccermendes},  and     the
structure, speed and play patterns of World Cup soccer final games between 1966-2010   \cite{JSMSEvolutionWorldCupfinal}.

These interesting papers are  at the interfaces of various disciplines  and are often tied to various technical questions or are limited to the analysis of distribution functions,  as often found for complex systems, but without conveying questions tied to self-organizations \cite{soc} or external constraints \cite{roehner2007driving}.

\section{UEFA team analyzed data sets}\label{UEFAdata}

\begin{table} \begin{center} 
\begin{tabular}[t]{lcccclclcccc} 
\hline   
  rank &  Team   name  &     Coeff.   &   rank & Team   name  &   Coeff.   
\\   
\hline 
 \;1 & FC Barcelona	& 134.6906&
 \;2 &   FC Bayern M\"unchen& 	114.8653\\ 
\;3 &  Manchester United& 	113.9074& 
 \;4 &   Chelsea FC	& 111.9074\\
 \;5 &  Real Madrid CF& 	111.6906& 
\;6 &   Arsenal FC &\;\;96.9074\\
 \;7 &  FC Internazionale &\;\;91.3962& 
 \;8 &   Atletico de Madrid &\;\;88.6906\\ 
 \;9 &  FC Porto	 &\;\;87.0668& 
 10 &   Valencia CF	 &\;\;84.6906\\ 
 \hline
 11 &  Olympique Lyonnais& \;\;84.0332&12 & FC Shakhtar Donetsk& \;\;80.6520\\ 
 13 & SL Benfica&\;\;79.1068&14 & Milan AC&\;\;78.3962\\
 15 & CSKA Moskva&\;\;76.4330&16 & Olympique de Marseille&\;\;75.0332\\
 17 &  Liverpool FC&\;\;68.9074& 18 & FC Schalke 04&\;\;66.8653\\ 
 19 &  Sporting Club Portugal	&\;\;66.0668& 20 &  Manchester City&\;\;64.9074\\ 
 \hline
  21 &  Villarreal CF&\;\;64.6906& 22 &  Dynamo Kiev &\;\;62.6520\\ 
 23 &  PSV Eindhoven &\;\;61.5461&24 &  FC Zenit St.Petersburg &\;\;60.4330\\
 25 &  Ajax Amsterdam &\;\;59.5461& 26 &  Sporting Braga	&\;\;59.1068\\
 27 &  SV Werder Bremen&\;\;57.8653& 28 & FC Twente Enschede&\;\;56.5461\\ 
 29 &  Metalist Kharkiv &\;\;53.1520& 30 &  Tottenham Hotspurs FC&\;\;52.9074\\ 
 \hline
31	&	Hamburger SV	&\;\;	52.8653	& 32	&	Sevilla CF	&\;\;52.1906	\\
33	&	Olympiakos Pireus FC	&\;\;51.4000	&
34	&	AS Roma	&\;\;50.8962	\\
35	&	VfB Stuttgart	&\;\;49.8653	&
36	&	Juventus FC	&\;\;49.3962	\\
37	&	Paris Saint Germain	&\;\;49.0332	&
38	&	Girondins de Bordeaux	&\;\;48.0332	\\
39	&	Athletic Bilbao	&\;\;47.6911	&
40	&	Standard de Li\`ege	&\;\;45.2000	\\
 \hline41	&	FC Basel	&\;\;44.5830	&
42	&	Bayer Leverkusen	&\;\;43.8657	\\
43	&	Fulham FC	&\;\;42.9074	&
44	&	FC Kobenhavn	&\;\;42.8600	\\
45	&	Lille OSC	&\;\;42.0332	&
46	&	Rubin Kazan	&\;\;40.9330	\\
47	&	Fiorentina AC	&\;\;40.3962	&
48	&	SSC Napoli	&\;\;40.3962	\\
49	&	RSC Anderlecht	&\;\;40.2000	&
50	&	Panathinaikos FC	&\;\;39.9000	\\
 \hline51	&	Udinese Calcio	&\;\;39.3962	&
52	&	AZ Alkmaar	&\;\;39.0456	\\
53	&	VfL Wolfsburg	&\;\;38.8653	&
54	&	Spartak Moskva FC	&\;\;37.4330	\\
55	&	APOEL Nicosie	&\;\;35.2170	&
56	&	Club Brugge KV	&\;\;35.2000	\\
57	&	Galatasaray SK	&\;\;35.1050	&
58	&	BATE Borisov	&\;\;33.9250	\\
59	&	Besiktas JK	&\;\;33.6050	&
60	&	Borussia Dortmund	&\;\;32.8653	\\
 \hline
\end{tabular} 
   \caption{ Top 60 UEFA ranked teams according to  the Sept. 2012 coefficients}
\end{center} \end{table}

Usually,  a ranking represents the overall performance over the period of a whole season. 
In particular, UEFA soccer teams are ranked according to results based on the five previous "seasons" for teams having participated in the  UEFA Champions League and  the UEFA Europa League\footnote{One should notice that it is not easy to  compare teams in different divisions or leagues.}.     The rules  leading to such a ranking are reviewed in App.A. They are more complicated than  a "win-draw-loss"  rating.  The ratings  depend on the success at some competition level, and differ according to the competition.  
 One should be aware that  a  
UEFA \underline{country coefficient} is used to pre-determine the number of  \underline{teams} participating for each
association either in the  UEFA Champions League  or in  the UEFA Europa League.
 
 A UEFA coefficient  table is freely available and is updated regularly depending on the competition timing. The present data, and its subsequent analysis, are based on the  Sept. 2012 downloaded Table.   
The team UEFA coefficients   are calculated as described in the  App. A boxes,  and are derived from

 $http://fr.uefa.com/memberassociations/uefarankings/club/index.html$. 
 
 The number of concerned  teams is  445; the UEFA coefficients range from $\sim$ 134.7 down to   $\sim$  0.383. For the sake of  illustration, the first 60 teams and their UEFA coefficient  (in Sept. 2012) are listed in  Table 1. 
 The statistical characteristics of these UEFA coefficients  are  given in Table 2,  2nd column.  Note  that  the kurtosis (a measure of the fourth moment of the distribution,  in fact equivalent to a specific heat in thermodynamics) changes sign  (near $r\simeq 35$)  according to the number of data points taken into account. 
 By analogy with phase transitions,  one should imagine  the existence of a "critical rank".  In fact,  the $\mu/\sigma$ function, which represents the order parameter in phase transition studies, evolves as  an exponential ($\sim e^{-r/73}$) rather than  as a   power law,   toward some  "critical rank".    If  the  analogy is  conserved,  such an  exponential behavior suggests  to consider the feature as one found at the  Kosterlitz-Thouless transition \cite{KT73}    in the 2D XY model, where  magnetic vortices are topologically stable configurations, - as in spin glasses or  thin disordered superconducting granular films.

 For completeness,   and noting that   the fits are non-linear,  the present study uses   the Levenberg-Marquardt algorithm  \cite{Levenberg44,Marquardt63,LMalgorithm,ranganathan2004levenberg} in  the present study, - except in Fig. 4 where a simple least square fit has been used  for simplicity.  The error characteristics  from the fit regressions, i.e. $\chi^2$, d, the number of degrees of freedom, the $p-value$ \cite{NJP14.12093038scoreprize,AoyamaTakayasu,gravetter2013essentials}, and the $R^2$ regression coefficient, are given in Table \ref{TableR2chi2dp}.  It can  be observed that in all cases the $p-value$ is lower than $10^{-6}$ (abbreviated by 0 in the Table).  Each $\chi^2$ value is rounded  to the closest  integer.


 \subsection{Empirical Ranking Laws}\label{rankinglaws}
Beside the classical  two-parameter power law, Eq.(\ref{Zipfeq}), 
  other,  often used,  three-parameter  statistical distributions,  can be used
  \begin{itemize}
  \item the  Zipf-Mandelbrot-Pareto  (ZMP) law  \cite{FAIRTHORNE}:
 \begin{equation} \label{ZMeq3}
y(r)=b/(\nu+r)^{\zeta}.
\end{equation}
\item
  the power law with exponential cut-off  \cite{Pwco3}:
  \begin{equation} \label{PWLwithcutoff}
 y(r)= c \;r^{-\mu} \; e^{-\lambda r}, 
\end{equation}
  \item the mere exponential  (two-parameter fit) case  
  \begin{equation} \label{expL}
 y(r)= d\;e^{-\eta r}.
\end{equation}
\end{itemize} 

  The ZMP law  leads to a 
curvature at low $r$  in a log-log plot and presents an asymptotic power
 law behavior  at large $r$.  
Note that both $\alpha$ and $ \zeta$ exponents, in  Eq.(\ref{Zipfeq}), and  Eq.(\ref{ZMeq3}). must be greater than 1 for the distributions to be well-defined (also greater than 2 for the mean to be finite, and  greater than 3 for the variance to be finite). 
  On the other hand, since $\nu$ in  Eq.(\ref{ZMeq3})  is not necessarily found to be an integer in a fitting procedure, $r$    can be considered  as a continuous variable,  for mathematical convenience, without any loss of mathematical  rigor; the same holds true  for the fit parameters $a$, $b$, and $c$, and   the "relaxation ranks" $\lambda$ and $\eta$.

 \subsection{Data Analysis}\label{dataanalysis}

\begin{figure}
 \includegraphics [height=13.0cm,width=13.0cm]{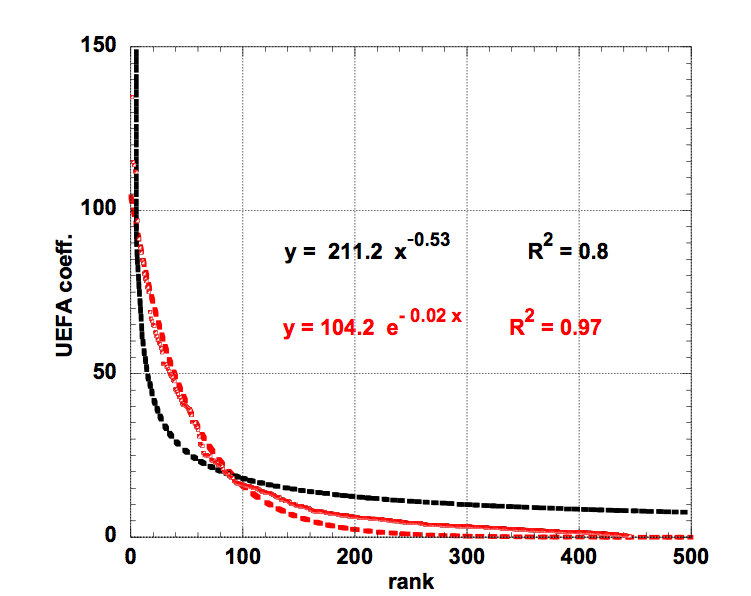}
 \caption   {  Possible empirical relationships between  their  UEFA coefficient (red dots) and the 445 team ranking in Sept. 2012;  both an exponential  (red dashes) and a power law (black dots)  fit are shown,  with the numerical  values of the parameter fits and  the corresponding regression coefficient $R^2$}    \label{Fig1Plot2colallteams2fitsa}
 \end{figure}

\begin{figure}
 \includegraphics [height=16.0cm,width=14.0cm]{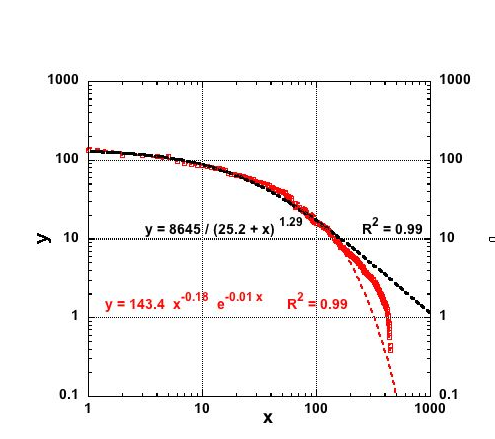}
  \caption   {  Possible empirical relationships between  their  UEFA coefficient  (red dots)  (here called  y) and the 445 team rank (here called  x), in Sept. 2012; the numerical  values of the  fitting  parameters and  the   regression coefficient $R^2$ corresponding to   a ZMP law, Eq.(\ref{ZMeq3}),  (black dots)  and   a power law with exponential  cut-off, Eq.(\ref{PWLwithcutoff}),   (red dashes) are given } 
 \label{Plot10UEFApwlctoflolof}
  \end{figure}

\begin{figure}
\centering    \includegraphics [height=12.0cm,width=13.0cm]{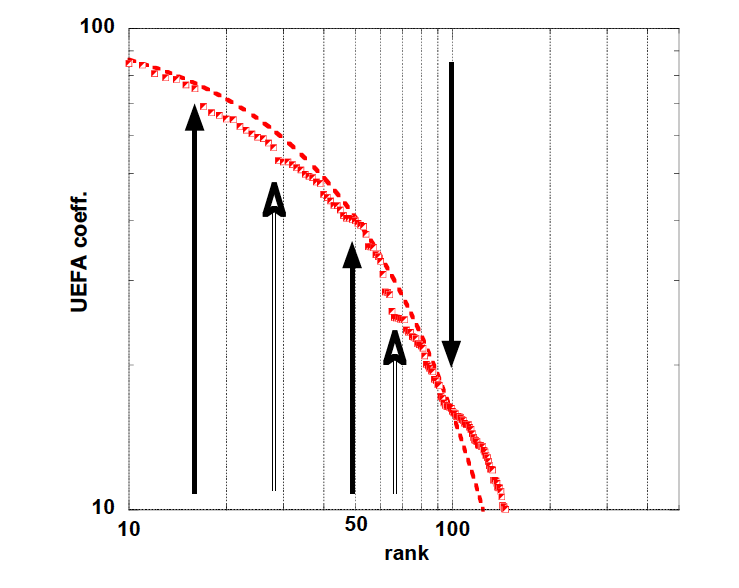}
 \caption   { Log-log plot of the   UEFA coefficient  values   (red dots) of   the (top 150) teams  in Sept. 2012.  Jump deviations from  the  power law with  exponential cut-off  (red dash lines), Eq.(\ref{PWLwithcutoff}), are indicated  by up pointing arrows at   various ranks between 10 and 100. The pointing down arrow marks the rank ($r\sim100$) separating the glass state from the disordered state   } 
\label{Plot10UEFAloloblowupexpf}  \end{figure}

A few simple and  possible various rank-size  empirical distributions  are shown Figs.\ref{Fig1Plot2colallteams2fitsa}-\ref {Plot10UEFAloloblowupexpf}.  

From Fig.\ref{Fig1Plot2colallteams2fitsa},  the exponential law  would appear to be more appropriate than the     power law, - in view of the regression coefficient $R^2$ values ($\sim 0.97$ $vs.$ $\sim 0.80$). However,  the origin of the numerical value of the coefficient in the exponential can be hardly imagined from   theoretical  arguments.  One can merely attribute it to some "relaxation rank" $\sim 50$.  The power law exponent $\alpha\sim 0.54$ on the other hand is  rather far from 1, and low\footnote{Note that the amplitude of the power law is about $r_M/2$} . In fact the marked deviations at low and high rank $r$, from the empirical fits, in particular from the single power law fit,  indicates the existence of different regimes.  
Note  the accumulation of data points $below$  the (power law)  fit occurring for  teams with very low $r$, i.e. $r\le5$. This suggests the existence of  a so-called queen effect \cite{Sofia3a} for the   top 5 teams. 

This   effect can be emphasized through the use of the ZMP law, Eq.(\ref{ZMeq3}), as shown in Fig.\ref{Plot10UEFApwlctoflolof}, - using a log-log plot for emphasis of the goodness of fit in the low rank regime.  Note the successive deviations of the data from the fits,   whence  again suggesting different regimes.
 In Fig.\ref{Plot10UEFApwlctoflolof},   a  fit by a  power law  with exponential  cut-off, Eq.(\ref{PWLwithcutoff}), is also displayed.  
 These three-parameter ZMP,  Eq.(\ref{ZMeq3}),  and    the power law with cut-off, Eq.(\ref{PWLwithcutoff}),  necessarily lead to a better  R$^2$ ($\sim0.99$) than the two-parameter fits.  A goodness-of-fit test indicates that the latter two empirical laws can be further considered.  The meaning of the $\nu$ value,  in the ZMP law Eq.(\ref{ZMeq3}),  has been discussed elsewhere \cite{Sofia3a,MABSS,[HB],[JM],[GR]}. On the other hand, the power  law with exponential cut-off   \cite{Pwco3} has been discussed  as  occurring from the "random group formation",  - in a sport research context \cite{NJP14.12093038scoreprize}. 
  
Thus, it can be admitted that marked deviations  also   occur for $r\ge 100\sim120$.  Therefore,  when examining the $r\le100$ range, several regular size regimes appear,   after  zooming on   the vicinity of the marked fit deviations at  ranks between 10 and 100 as shown for  the  power law with cut-off case, on a log-log plot,    in Fig.\ref{Plot10UEFAloloblowupexpf}.  Successive arrows indicate "data  steps" at $r\simeq$ 16, 28, (39), 50, 62, 73, at least. This allows us to emphasize an intrinsic structure in such  regimes with a "periodicity" $\simeq$ 11 or 12.

   \begin{figure}
\centering  \includegraphics [height=14.0cm,width=14.0cm]{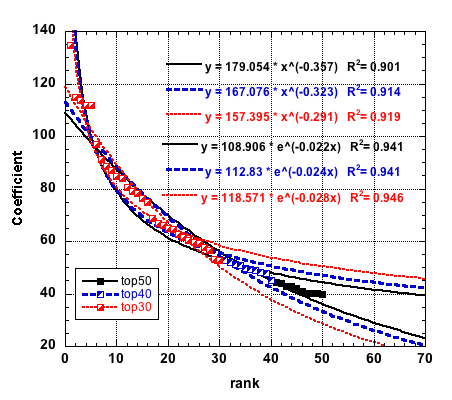}
\caption   {  Possible empirical relationships between  the  UEFA coefficient for the top  30 (red half filled squares), top 40  (blue half filled squares), or top 50  (black filled squares),  teams, as ranked in Sept. 2012. Both an exponential and a power law fit are shown,  in each case, with colors corresponding to the respective data. The numerical  values of the parameter fits and  the corresponding regression coefficient $R^2$ are given  } 
 \label{Fig2Plot2foottops} 
\end{figure}

   \begin{figure}
\centering  \includegraphics [height=14.0cm,width=14.0cm]{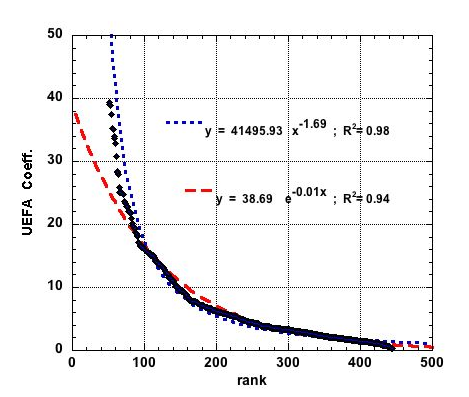}
\caption   { Possible empirical  relationships between  the  UEFA coefficient  (black diamonds) for  teams  above  the  50 rank  and their rank  in Sept. 2012. Both an exponential  (red dashes) and a power law (blue dots)  fit are shown,  with the numerical  values of the  fitting parameter  and  the corresponding regression coefficient $R^2$  } 
 \label{Fig3Plot3teamsbelow502fits} 
\end{figure}


Therefore, since different regimes  can be seen emerging  at various intervals, the  team ranking behavior can be more precisely  re-examined. This leads to  much  statistical analysis and many  fit trials. 
Two types, either an exponential or a power law,  are shown for three different sub-selections in Fig.\ref{Fig2Plot2foottops}, i.e. the top 50, 40 and 30 teams.   The statistical characteristics of the "sample" distributions are given in Table 2,  columns 3-5. The increase in the $R^2$ value with respect to the overall regime (in Fig.\ref{Fig1Plot2colallteams2fitsa})  is remarkable, for the power law fits, i.e. $R^2\simeq0.797\rightarrow 0.98$, suggesting that the  50 top teams or so    "behave"  in a different way  from the others $r>50$.Observe the value of the power law exponent: $\alpha \sim 0.3$, in Fig. 4,  in this regime, instead of 0.53 for the whole set, in Fig. 1. The evolution of the regression coefficient is   very mild when changing the "sample size", see Table 3. Observe that the
 fits in Fig. \ref{Fig2Plot2foottops}  do not indicate any striking difference between the exponential and power law fits, from the $R^2$ or $\chi^2$ value criteria.  Note  from Fig. 4 that the numerical value of the exponential  "relaxation rank" , i.e. the prefactor $ (\simeq0.02)$ for  $-x$ , is  still $50$, like the value of the "relaxation rank"   in Fig.\ref{Fig1Plot2colallteams2fitsa} .

Next, the behavior of the teams  above $r=50$ can be quickly examined for completeness: see Fig. \ref{Fig3Plot3teamsbelow502fits}. Observe  that  the exponent $\alpha $  $\sim 5/3$ for these high ranking teams differs  significantly   from the corresponding one for  (the best) low ranked teams, $\sim 0.36$, as displayed in Fig. 4.  Again, this   emphasizes  some difference in "behavior" between the teams ranked below or above $r\simeq$ 50.   The method of "primacy analysis"  seems  thus  of  subsequent interest.

   \begin{table} \begin{center} 
\begin{tabular}[t]{cccccccccc} 
  \hline 
   $ $   &UEFA coeff &top50 & top40&top30&below50    \\ 
\hline   	

\hline  
min&  0.3832&	39.9&45.2&52.907&  0.3832  \\
Max&	 134.69&134.69&134.69&134.69& 39.396 \\
mean ($\mu$)&	14.053&	65.586&71.531&78.822&7.5299 \\
Median &	5.475& 59.326&64.799&75.733&4.4250\\
RMS &	25.45&69.553&74.896&81.582&  10.832 \\
Std. Dev. ($\sigma$) &21.242&23.389&22.482&21.402&7.7970\\
Variance &451.23&547.03&505.42&458.06&	60.794  \\
Std. Err. &	1.007&3.3077&3.5546&	0.474&0.3923  \\
Skewness  &2.6887&1.0773&1.0167&0.89635&1.8917 \\
Kurtosis &7.8008&	0.45378&0.2470&-0.02845&3.5043\\
 $\mu/\sigma$ &0.662&2.804 &3.181 &3.683 & 0.9657  \\
  \hline
\end{tabular} 
   \caption{Summary of  statistical  characteristics  for  the Sept. 2012 UEFA coefficient team ranking data   }\label{TablestatUEFA}
\end{center} \end{table} 

         \begin{table} \begin{center} 
\begin{tabular}[t]{cccccccccc} 
  \hline 
   $ empirical$ $law$&Eq.&Fig.   &$\chi^2$&d &  p &$R^2$ \\ 
\hline   	
\hline  
pw& (\ref{Zipfeq})&	1&40678&444&  0&0.797  \\
exp&	  (\ref{expL})&1&5826&444& 0&0.972 \\\hline
pwco&(\ref{PWLwithcutoff})&2,3&1477&444&0&0.99 \\
ZMP&(\ref{ZMeq3})&2&2659&444& 0&0.99\\\hline
pw &	(\ref{Zipfeq})&4&2195 &49&   0&0.92 \\
pw&(\ref{Zipfeq})&4&1474&39&0&0.93\\
pw&(\ref{Zipfeq})&4&975&29&0&0.93  \\
exp& (\ref{expL})&4 &1417&49&0&0.95  \\
exp& (\ref{expL})&4&1076&39&0&0.95 \\
exp& (\ref{expL})&4&677&29&0&0.95\\\hline
pw&(\ref{Zipfeq})&5&216&394&0&0.98  \\
exp&	(\ref{Zipfeq})&5&821&394&0&0.94 \\\hline
  \hline
\end{tabular} 
   \caption{Summary of   regression fit characteristics: $\chi^2$, d: number of degrees of freedom; p : $p-value$, and   $R^2$: regression coefficient \cite{gravetter2013essentials}
 }\label{TableR2chi2dp}
\end{center} \end{table} 

\section{Analysis of primacy}\label{primacy}

 It has been seen here above  that the   UEFA  team ranking distribution 
can be close to a  rank-size relationship. However, 
these distributions are primate  distributions \cite{s1}, i.e. 
  one or  very few    teams 
predominate  the distribution shape leading to a   
convex  distribution  that corresponds to the presence of  a   number of teams, 50 or so,  with  much larger  coefficients than the mean coefficient  $\sim 14$  (std. dev. $\sim21.25$; see Table \ref{TablestatUEFA}). Therefore, concentrating on such "top teams", it is of interest  to raise the question whether the UEFA coefficient ranking "method" implicitly induces some inner structure. In order to do so, the notion of primacy measure is developed here below.

\par \subsection{ Sheppard Primacy measure} \label{Primacy measure}
Measures of "primacy" can be of the kind

\begin{equation} \label{a4}
Pr_{(k-1)}^{(1)} = \frac{N_1}{\sum_{r=2}^k N_r},   \; \mathrm{with } \;   k=2,3,\dots ,
\end{equation}
 measuring the percentage of a contribution in the whole distribution. In particular,  Eq.(\ref{a4}) gives a numerical value for the  primacy of the best ranked entity with respect to the next $k-1$  entities, since these are ordered by decreasing
values.

One can go on and define 
\begin{equation} \label{a4ma}
Pr_{(k_M-1)}^{(j)} = \frac{N_j}{\sum_{r=j+1}^{k_M} N_r}, \; \mathrm{with } \;  k_M=2,3,\dots ,  
\end{equation}
in order to measure the  primacy of  any entity $j$ over  selected $k_M-j-1$ lower entities. Obviously this number is reduced  when $j$ increases, due to the necessarily finite size  $ k_M \le r_M$ of the system.

 If a power law of the kind $N_r =  N_1/r^\alpha$, - see  Eq.(\ref{Zipfeq}),   is substituted into
each of these measures, it is obvious  that the corresponding  index of primacy depends on $\alpha$. Whence 
 rank-size relationships with different  $\alpha$ will have 
different "levels of primacy", which consequently will be hardly comparable to each other.
 
Sheppard \cite{s1} tried to avoid this puzzle
by formulating  a {\it primacy index} that is independent of $\alpha$ 
(see Appendix A for some further introduction).  He defined
\begin{equation} \label{a7}
Pr_N =\frac{1}{N-2} \sum_{r=1}^{N-2} \bigg[ \frac{\ln(N_r+1)-\ln(N_{r})}{\ln(N_{r+2}) - \ln(N_{r+1}))} \bigg] 
\bigg[ \frac{\ln(r+2)-\ln(r+1)}{\ln(r+1)-\ln(r)} \bigg].
\end{equation}

\par \subsection{Modified Sheppard-idea based primacy indices } \label{NewPI} 
However the primacy index of Sheppard, Eq.(\ref{a7}), contains the  difference between two logarithms in the
denominator. Thus, when  two  consecutive UEFA coefficients  have almost the same  value, 
 this  difference can be very small,  thus leading to a huge value of
the Sheppard index.   In order to avoid 
such  problems,    other "local primacy measures"   can be considered,  forcing the  difference between two closely related (logarithms of ) $N_r$ 
 to be present only in the numerator. 
 Keeping  $N_{r+1} \le N_{r}$, (and $r+1>r $), these measures are 
\begin{equation} \label{v1}
V_{r}=\;-\;\frac{\ln(N_{r+1})-\ln(N_r)}{\ln(r+1) -  \ln (r)}
\end{equation}
and 
\begin{equation}\label{v2}
W_{r} =   \frac{\ln(N_r) - \ln(N_{r+1})}{
\ln(r+1) - \ln(r)} -  \frac{\ln(N_{r+1}) - \ln(N_{r+2})}{
\ln(r+2) - \ln(r+1)} \equiv V_{r} - V_{r+1}
\end{equation} 

\begin{figure}
\centering
 \includegraphics [height=13.0cm,width=13.0cm]{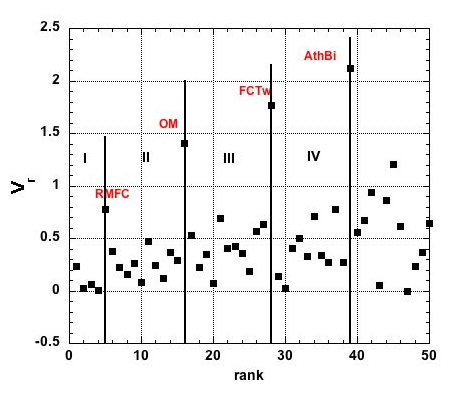}
\caption   {   Primacy index $V_r$  indicating some inner structure in the  50 top  teams  ranking, in Sept. 2012. The "border lines" are emphasized by team acronyms} 
 \label{Fig4Plot5Vrtop50} 
\end{figure}

\begin{figure}
\centering
 \includegraphics [height=13.0cm,width=13.0cm]{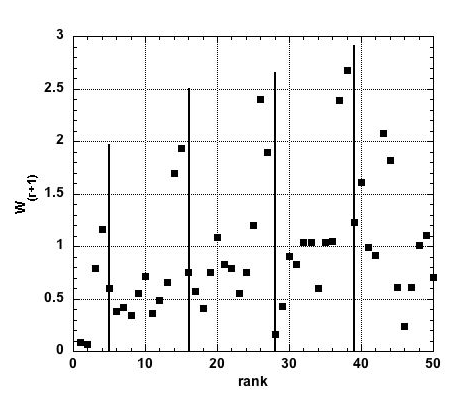}
\caption   {  Primacy index $W_{r+1}$  reflecting the  inner structure, found in the $V_r$ $vs.$  $r$ plot,  in the  50 top UEFA  teams  ranking, in Sept. 2012. The "border lines"   define "classes",  as emphasized by the sharp  $W_{r+1}$  transitions  } 
 \label{Fig5Plot25Wr1top50} 
\end{figure}

 Observe that $V_r$  and $W_r$ are  related to the next team(s) in the ranking, and measure a sort of distance.
 The results of the applications of such formulae for the problem at hand  are shown in Fig. \ref{Fig4Plot5Vrtop50}  for  $V_r$ and Fig. \ref{Fig5Plot25Wr1top50} for $W_{r+1}$. 
 
 These enlightening figures indicate  well defined  regime borders, corresponding to specific teams, as indicated in Fig. \ref{Fig4Plot5Vrtop50}.  For $r\ge5$,  the method confirms that the successive "regime sizes"  remain rather equal in value ($\sim 11$ or 12). There is also some "accumulation" of teams near the border, e.g. as seen in Fig. \ref{Fig5Plot25Wr1top50}, - reminding of a sharp transition behavior.    
 
These features are   also reminiscent  of the occurrence of some sort of  granular structure along a 1-dimensional space, or more generally dissipative structures  \cite{dissip}. Several analogies come in mind, e.g.: (i) hot spots in ballast resistors \cite{BDMaz47,BDMaz46} or  superconductors \cite{PhB108hotspot};    (ii)   invasion waves in   combustion   \cite{[9],[10]},    (iii) Marangoni tears, and  (iv) Benard cells,  as in cloud band structures \cite{tellus11.59.267.kuettnercloudbands}, or (Spitzberg) stone structures \cite{romanovsky1940application,romanovsky1941application}.
 In all such so called intermittent processes  in an open medium, the interplay between feedback mechanisms  is known to induce temporary  structures.\footnote{The connection between transitive continuous ordering of states by a ranking function and thermodynamic entropy is  known in conventional thermodynamics under the name of adiabatic accessibility. This principle was introduced by Caratheodory \cite{[24]} and  used by Lieb and Yngvason \cite{[25]}
See physical arguments in favor of connecting the entropy potential to the rank in
 \cite{[9],[10]}.}  

In order to illustrate such features through a simple model, consider the
nonlinear open system known as the 1-dimensional ballast resistor.
 It describes the interplay between the input current leading to a Joule effect ($RI^2$)  and the energy dissipation due to the heat transfer with the
 surrounding environment wherein the  transfer is controlled
  by the heat capacity and heat conductivity of the wire \cite{BDMaz47,BDMaz46}.  In a first approximation,  this leads to searching for the stability points of a system in a cubic potential. Whatever the boundary conditions, the system can become heterogeneous with hot and cold domains which themselves can become heterogeneous through Hopf bifurcations \cite{PhB108hotspot}. The sizes of the domain and their motion (and life time) can be calculated \cite{BDMaz47,BDMaz46,PhB108hotspot}. The connection to an open\footnote{the number of  UEFA ranked teams changes every year or so}   social system,   for which the internal ranking rules look like unstable conditions for self-organization, is not immediate. However,  the analogy  seems pertinent enough to   suggest  that models of intermittent processes can  serve as  bases for further generalization.

For  illustrative purpose, a (very imperfect) toy model is presented in App. C in order to show that  UEFA like rules, recalled in  App.A,  i.e. attributing points to teams in a (necessarily) biased way over several seasons, imply the opening of gaps in the UEFA coefficient structure, thus regimes.  The model is  quite simplified, whence a search for properties similar to the mentioned Kosterlitz-Thouless transitions or  characteristics of dissipative structures is quite outside the scope of the present paper.

\section{Discussion}

As seen in  Fig. \ref{Fig3Plot3teamsbelow502fits},  the exponent $\beta $ $\sim 5/3$  for the high rank team regime much differs from the corresponding one $\sim 1/3$ for  (the best) low rank  teams (Fig.4).
In a scientific field, like  condensed matter physics,  where exponents are  frequently  used for sorting out processes, a   low value of the exponent stresses the narrow range of "activity".  The five-fold  ratio   between two exponents would indicate  somewhat "long range" or "diffuse properties"  for the system  (= regime) with the largest exponent.

As in thermodynamics, the  $\beta $ exponent  drastically different values may indicate a "state phase" change,  between a compressed phase and one (or more)    disordered phases. In fact, one can further analyze the data from a statistical point of view, arguing that the ratio $\mu/\sigma$ between the mean and the standard deviation  of rank-size distributions can be considered as the order parameter, as  in studies of phase transitions. As shown in Table 2, such a ratio steadily increases when reducing the considered number of  "top of the top" teams, - forming the ordered state. On the other hand, the ratio $\mu/\sigma$  decreases  below 1 for the  "below  the top 50" teams, thereby indicating a sort of  ordered-glass state phase transition.  The pointing down arrow in Fig. 3  marks  such a "critical rank"  between the medium rank teams (glass or vortex state) and the high rank teams (disordered state), in the present case.  
 
Such findings on structures in  organized competitions,  like the UEFA  Champions League and Europa League, invite some thought. In these   apparently specific  (team) competitions,  the "self-organization scores" determine that the predominant direction of evolution of the system is directed towards regular structures, thus $less$ $opened$ competitiveness. These competitions seem to be paradigmatic cases. However, the findings suggest ways  for discussing many other, more or less abstract, competitions,  as  complex  social systems are, when there are imposed competition rules (and scores) in their generic form \cite{[9],[10]}. This  induced structuralization was also seen in other (peer-to-peer) competitions between industrial companies  \cite{caram2010dynamic}.   Therefore,  one can audaciously state that ranks in social competition might be unavoidable  and  structures can hardly be modified,  - because of classical thermodynamics principles. It remains to be studied and discussed whether quantum-like  ("probabilistic") constraints would be allowing more competitiveness and subsequent   ranking modification.




\section{Concluding remarks}\label{conclusions}

First, it  should be  stressed that   it has been searched whether the classical rank-size relationships \cite{newman,Brakman}  can be treated as indicators of a {\it  sport  team class system}, taken as a complex  system and worthy of a scientific approach, - as are many dynamical systems nowadays.  A norm of  performance for  teams  or    team budget effects  have  not  been considered. In this paper,  we  discuss the   ranking in the  particular case of   UEFA teams,  at first  not debating the value of the points attributed to victory or loss at some competition level.  Yet, their somewhat relative values arising from some arbitrariness  underlines   the present  investigation and leads to harsh conclusions.

First, a  conclusion  is reached that the distribution of  UEFA team ranking does $not$ follow   a single power law  nor an exponential.  Instead, it appears that the  ranking should be  grouped into  classes. The rank-size  distribution for  these classes can be  approximated by a mere  scaling (power) law,  with a  quite different exponent,  $\simeq$ 1/3 or 5/3,  - suggesting a sort of order-disorder phase transition, in a thermodynamics-like sense, between the  lowest and highest  rank teams.  Moreover, through this log-log search for an empirical law, it is found that structures exist in the overall classification. The medium rank regime is made of several "glassy" states. The very high ($r\ge100$) ranks corresponding to a disordered state. Along this analogy, the rules establishing the team UEFA  coefficients act like a magnetic field   inducing various structures in a magnetic or superconducting system.

 In addition,  in Sect.\ref{primacy}, an index has been introduced in order to enhance  regimes in data ranking.   Through the "team primacy"   investigation for the top group of    teams,  on the basis of  a modification of the conventional  Sheppard  index of primacy, the  regimes are emphasized as  extrinsically  induced  structures.
It has been indicated that  measures, called $V_r$ and $W_r$,  independent  of the empirical law functional form,  lead  to subsequent numerical results pointing to  some inner structure of the team rank-value distribution,   according to the present set of  UEFA rules.   It can be guessed, without  more mathematical work than  is necessary, that the inner structure is likely tied to the pool, in the first rounds,  followed by the  "direct elimination" tournament-like process  \cite{EPJB75.10.327soccermendes} in UEFA competition, and the "increase in value"  of the team when progressing further in the competitions through an accelerating effect. with  a set of nonlinear constraints (similar to boundary conditions) inducing  an unstable organization,  similar to what is found in intermittent processes, when forced cycles drive an interplay between (two or several) feedback mechanisms and leading to visible physically characterized zones \cite{dissip}.

The  (very imperfect) toy model presented in App. C  points to  the opening of gaps when values of  game results and tournament forms are imposed.     Nevertheless, this   very simplified model should suggest to the reader  further research, through simulation or analytic  work, in order   to study equivalent  properties similar to those characterizing the mentioned Kosterlitz-Thouless transitions or  characteristics of dissipative structures, hinted to be appearing in the UEFA coefficient rank-size structure, thus  biased regimes.

 In general, strongly intransitive competitions, in any society, - dare we   say,  can display types of behavior associated with complexity, like competitive co-operation and leaping cycles.  External influences induce structures in many physical processes. Through an example, we bring some evidence of the universality of the  features. Note that the analogy with strict  thermodynamics weakens as competitive systems become more intransitive  \cite{[9],[10]}.

In future work, it would be of interest to examine whether the  Sheppard generalized  ideas and measures  could be used to quantify  how much a complex system  rank-size distribution deviates   from the usual power-law rank-size relationship, Eq.(\ref{Zipfeq}).  Moreover, one could  ask whether 
  this  hyperbolic form  has truly to be  chosen as an optimum basis for discussing rational ranking rules. 
These measures  seem to  allow one  to introduce a finer description of   classes. In so doing, they might also  serve  for weighting performances,  as in  the case of NCAA College Football Rankings  \cite{JSE10.09.582weightranking},    by organizing  more homogeneously (depending on team budgets and expectations  \cite{NJP14.12093038scoreprize}) based team competitions or regulating  various  sport conditions in  team  ranking.  This may also imply some  consideration  to relaxing  the rule about the number of teams of a "country" which may play in the UEFA competitions,   if more competitiveness is of interest. An impossible dream?   
 
Sports other than soccer are  open to further investigations. Then, it  will be interesting to see whether some universal  features occur or  whether there are marked  differences.

\bigskip
\begin{flushleft}
{\bf Acknowledgments} 
\end{flushleft}
This work has been performed in the framework of COST Action IS1104 
"The EU in the new economic complex geography: models, tools and policy evaluation".
MA and NKV acknowledge some support
through the project 'Evolution spatiale et temporelle d'infrastructures r\'egionales 
et \' economiques en Bulgarie et en F{\'e}d{\'e}ration Wallonie-Bruxelles', within the  
intergovernemental agreement for cooperation between the Republic of  Bulgaria and  the  Communaut\' e  Fran\c{c}aise de Belgique. A partial support by BS39/2014 is acknowledged by AG.  Thanks to Peter Richmond and  Wayne K. Aug\'e II   for much  improving  the readability of the paper.
 
\newpage
 
{\bf Appendix  A:  UEFA score rules  }

 \vskip0.4cm
\fbox{%
\begin{minipage}{5 in}

{\bf Box \#1: Champions League points system}
\vskip0.4cm
 
\begin{itemize} \item 1st qualif. round elimination : 0.5 pts, \item 2nd qualif. round elimination : 1 point,\item Group stage participation : 4 points, \item Group stage game win : 2 points,  \item Group stage game draw : 1 point, \item Round of 16 participation : 4 points
\end{itemize}

N.B. 

Since the 2009/2010 season, clubs have been awarded an additional point if they reach the round of 16, quarter-finals, semi-finals or final.

Points are not awarded for elimination in the 3rd qualifying round or play-offs because  the teams go to Europa League 
\end{minipage}}

\vskip 0.4cm
\bigskip

\fbox{%
\begin{minipage}{5 in}

{\bf  Box \#2: Europa League points system}
\vskip0.4cm

\begin{itemize} \item 1st qualif.round elimination : 0.25 pts
 \item 2nd qualif. round elimination : 0.5 pts,
 \item 3rd qualif. round elimination : 1 point,
  \item Play-off elimination : 1.5 points,
  \item Group stage  game win : 2 points,
    \item Group stage game draw : 1 point,  
\end{itemize}  
 
N.B. 

Since  the 2009/2010 season, clubs have been guaranteed a minimum of 2 points if they reach the group stage,  and are awarded 1 additional point if they get to the quarter-finals, semi-finals or final.
\end{minipage}}
 \vskip0.5cm

Some  brief explanation, e.g. based on the 2013-14 UEFA competitions,  can be useful in order to follow the competition organization,  team elimination rule and subsequent point counting,  as indicated in the boxes here above.  How many teams are selected, i.e.   qualified, for the UEFA Champions League 	and the UEFA Europa League,  is somewhat irrelevant for the present purpose. This is not discussed further. Details can be found in e.g.  
$http://en.wikipedia.org/wiki/2013-14_{-}UEFA_{-}Champions_{-}League$
and  in
$http://en.wikipedia.org/wiki/2013-14_{-}UEFA_{-}Europa_{-}League$.   

A total of 76 teams  
participated in the 2013-14 UEFA Champions League.  Such a number of teams and the subsequent number of  rounds depend on the number of available days for such a competition  in a season.  First, note that the qualif. round and play-off eliminations  are direct confrontations between 2 teams, somewhat drawn at random.  The process went as follows in 2013-14:   four less well ranked teams  had to go to the so called 1st qualif. round elimination. The two  losers got each 0.5 pts; the winners joined 32 teams at the  2nd qualif. round elimination.  The 17 losers got 1 pt each.  The 17 winners of the  direct confrontations qualified for the 3rd qualif. round elimination with 13  other teams.  The \underline{15 losers} went to the Europa League  play-off round (see below).    The 15 winners of  direct confrontations qualified for the Play-off round with 5 other teams.   The \underline{10 losers} went to the Europa League Group stage round (see below). The 10 winners joined 22 "directly best qualified teams" for the "Group stage". These 32 teams were were drawn into eight groups of four. In each group, teams played against each other (home-and-away) in a round-robin format. The group winners and runners-up advanced to the round of 16, while the \underline{(8) third-placed} teams  were directly entered in the 2013-14 UEFA Europa League round of 32 (see below). Thereafter, teams played against each other over two legs on a home-and-away basis,    in a direct knockout phase,  up to the \underline{one-match final}. For the group stage and later rounds, the points were attributed as mentioned in Box \#1.
 
In contrast,   194 teams  participated in the 2013-14 UEFA Europa League.  Among the 194 teams,  33 (=15+10+8; see above) were transferred from the Champions' League, according to the following process. The 161 qualified teams  were ranked (from bottom to top) and inserted into the qualifying founds as follows. 
In the first qualifying round, the 76 (less well ranked) teams played at home and away after a mere drawing (with the restriction that two teams from the same association could not play against each other).  The 38 winners were merged with 42 better ranked teams for the 2nd qualifying round. The 40 winners went to the 3rd qualifying round and merged with 18 better ranked teams.  The  
    29 winners from the third qualifying round were merged with  better 18 ranked teams  AND 15 losers (see above) from the Champions League third qualifying round   to make a "play-off round" of 62 teams. The 31 winners from the play-off round went to the  Group stage made of  48 teams, i.e. 7 directly qualified and 10 qualified from the play-off round of the Champions League (see above).   	
 The  24 winners were merged with  the 8 "third placed teams from the Champions League Group stage" (see above) to form the  "Round of 32". A direct knockout phase goes on thereafter on a home-and-away basis legs,  up to the single match final. The 38 losers in the 1st qualif. round got 0.25 pt each; see Box \#2.  It is easy to note the tam points thereafter.

 In so doing,  it is  easily noted that the teams get points according to the  level at which they were eliminated. Moreover,  the number of obtained points    increases much  when  a team participates in  games at the group stage level. 

 \vskip0.4cm
 
  \bigskip  \bigskip
  {\bf Appendix  B:  on Sheppard Index of Primacy }
  \bigskip    
  
Sheppard \cite{s1}  proposed the following index of primacy;

\begin{eqnarray}\label{a7c}
Pr_N=\frac{1}{N-2} \sum_{r=1}^{N-2} \bigg[ \frac{\ln(N_r)-\ln(N_{r+1})}{
\ln(N_{r+1}) - \ln(N_{r+2}))} \bigg] 
\bigg[ \frac{\ln(r+2)-\ln(r+1)}{\ln(r+1)-\ln(r)} \bigg].
\end{eqnarray}
Note  the following  logics behind  this index:   substitute   the power law
rank-size relationship, $N_r = N_1 r^{-\alpha}$, into the previous equation. The result is
\begin{equation}\label{a8c}
Pr_N=\frac{1}{N-2} \sum_{r=1}^{N-2} 1 = 1
\end{equation}
{\it whatever  the value of}  $\alpha$. Thus, for a  rank-size relationship obeying a perfect power law, the index $Pr_N$ has a value of
1, irrespective of the slope of the relationship on a log-log plot.

If    $Pr_N$  is less than 1, the value suggests  a locally convex form   for  the distribution.  This  suggests the occurrence of some  "strong primacy".  

However, if  (i) the  size histogram fit deviates from the $perfect$ power law,  like in the figures in the main text, and in fact  this is quite generally so\footnote{No need to emphasize that this is always the case if $R^2$ is strictly $< 1$},  and (ii)  if  the data is on the left  of the fitting curve,  then  $Pr_N$ will  necessarily exceed 1, as in the queen effect.   

In conclusion, the $V_r$ and $W_r$ measures  do well quantify the possible features.  The structural findings   are henceforth much enhanced.

  \vskip0.4cm
 \bigskip   
  {\bf Appendix  C:  Toy model}
  \bigskip

     In this Appendix, we want to indicate that the rule (see Appendix A) like {\it clubs have been awarded an additional point if they reach the round of 16, quarter-finals, semi-finals or final.} necessarily implies the opening of gaps in the ranking coefficient, whence the appearance of classes. In order to do so, we are much simplifying the UEFA competition conditions, and refer only  to  a virtual set of  top teams for  a  tournament. We let more complicated numerical   simulations for further work.
  
 The simplification is as follows: we take a "never any draw occurrence" hypothesis, and  consider the competition to be  limited to the round of 16 participation and upper levels. We have $not$ taken into account  that UEFA is, as emphasized here above, not only giving extra points  to a team going to the next round,  but also that such a number of points depends on the level of the competition.  This rewarding effect only enhances the number of points for the "best" teams, and  truly $stresses$ the point distribution in favor of such teams. We wish to make no "moral comment" on this point, but this reminds of a type of Matthew effect \cite{Merton}:  the winning teams are $quasi$ always  the same ones, and  stay more  at the top than others, though the ranks may change, of course.
 
 Grossly speaking, it can be admitted for simplicity that all the teams have many victories in the previous rounds, and have quasi the same number of points,  due to an approximate number of victories (2 pts), or of draws (1 pt). We are much aware that this is not the case  \cite{JAS36.09.soccer}, but the point distribution in  the first rounds, here serves only as a  rough background for the present considerations.  At the round of 16 participations, all the 16 teams already got 4 points, because they participate (see Table 1 in App. A). Thereafter, 8 teams, say  E, F, G, and H are eliminated, but have gotten 5 points because they participate at the  round of 8;  4 pursue to the next round, say A, B, C and D. Taking the most extreme case, i.e. all these teams won (and thus make no draw), they get 9 points. For each further level, under  the "never any draw" hypothesis, the winner at the final gets 24 points, the looser 20,  the semi finalists which did not move up get 14 points, etc.

\begin{figure}
\centering
 \includegraphics [height=13.0cm,width=13.0cm]{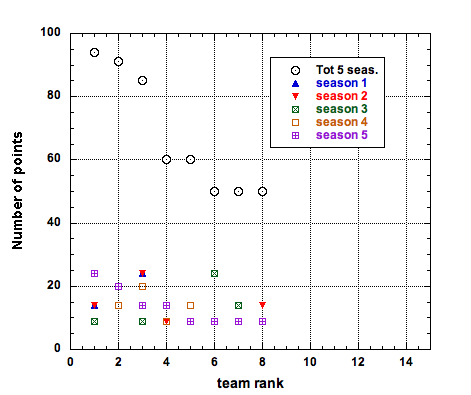}
\caption   { Simulated distribution of points, according to the  "toy model" in App. C, for a 5  season competition between 16 teams;. Only the top 8 results are given in each season together with the final result} 
 \label{Fig8Plot10simul5yrsAH} 
\end{figure}

 Consider 4 other seasons, and let the 8 top team be always the same ones.  However, randomize the distribution of points. After a finite number of simulations, 100 for the sake of this Appendix, and  after averaging the distributions, we obtain a distribution of points for the 8 top teams. The  final number of points  is rounded to  the nearest integer. Under such a (very simplified) tournament process, the final distribution is found to be that displayed in Fig. \ref{Fig8Plot10simul5yrsAH}.
 
Even though this is not  exactly equivalent to the   UEFA cases, the main emphasis on attributing rewarding  points after winning  some group and level stage is conserved. The equivalence  between reality and the toy model resides in the "legal decision" of  these attributed points through a ranking based on {\it a set of games} (irrespective of the detailed results). It should be obvious that the more group stages a team is wining,  the more so  the ranking level will be emphasized,  and accelerated depending on the level of the stage.  Exactly as in the toy example (no draw) tournament, the difference in the number of points between the best, medium, and worse teams will increase, and will be larger between the top teams (at low ranking) than for the high rank teams. Therefore, gaps will open up in the ranking coefficients, as already demonstrated in Fig. \ref{Fig8Plot10simul5yrsAH}.  It is (and in fact it will be) very difficult for teams to improve their ranking even though they  can  improve, - and conversely. Recall that the ranking is based on results over 5 seasons. Three recent cases are at hands for soccer fans or experts: Borussia Dortmund moved only to  the 15th place in 2014 from the 60th in 2012,  and Paris St Germain remained at the 37th place, while F.C. Barcelona is still at the second rank.
   
\begin{figure}
\centering
 \includegraphics [height=13.0cm,width=13.0cm]{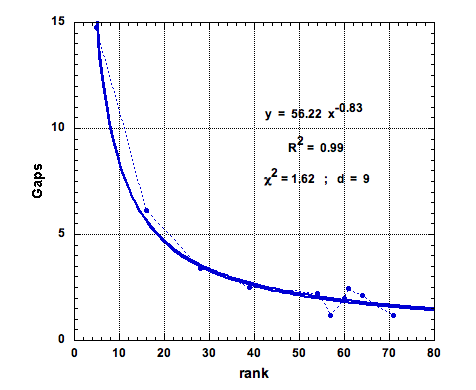}
\caption   {  Envelope function to the distribution of  the  10 first  gap  coefficient magnitudes  for the  UEFA ranking of the top   teams  in  Sept. 2012} 
 \label{Fig9Plot7envelopmaxderiv} 
\end{figure}

   As a final numerical feature, the envelope of the (absolute value) of the maxima of the derivative of the UEFA coefficients can be  precisely obtained through a hyperbolic fit, see Fig. \ref{Fig9Plot7envelopmaxderiv}, as $\simeq r^{-0.83}$, with $\chi^{2} \sim 1 .62$  for 9 degree of freedom, pointing to a  0.998 confidence level, - somewhat demonstrating the rigidity of the   gap structure.



\newpage
     \begin{thebibliography}{99}


\bibitem{compl1}
Axelrod R  and Cohen  M  D  1999  {\it Harnessing complexity: Organizational implications on a scientific frontier}  (New York, Free Press) 

\bibitem{compl2}
Oltivai Z  N  and  Barabasi  A  L  2002  {\it Life's complexity pyramid},   {\it Science}  {\bf 298},  763-764 

\bibitem{compl4}
Arthur W  B  1999  {\it Complexity and the economy}   {\it Science}  {\bf 284}, 107-109 

\bibitem{syst1}
Puu T  and Panchuk A  1991  {\it Nonlinear economic dynamics}  (Berlin, Springer) 

\bibitem{syst33}
 Bertuglia C S and Vaio F  2005   {\it  Nonlinearity, Chaos, and Complexity: The Dynamics of Natural and Social Systems} (Oxford University Press, Oxford)

\bibitem{kwapien2012physical}
  Kwapie{\'n} J  
  and Dro{\. z}d{\. z} S  
 2012  {\it Physical approach to complex systems},     {\it Phys. Rep.}  {\bf 515}, 115-226
 
\bibitem{tsallis2012entropy} Tsallis C 
2012  {\it  Entropy},   {\it Computational Complexity}    (Springer, New York)  pp.  940--964 

\bibitem{ts}
Thomson J  M  T and Stewart H  B  1986    {\it Nonlinear dynamics and chaos: Geometrical methods for scientists} (New York, Wiley)

\bibitem{str}
Strogatz S  2001   {\it Nonlinear dynamics and chaos: with applications to  physics, biology, chemistry and engineering}  (Reading, Ma, Addison-Wesley)

\bibitem{vesp}
Vespignani A  2009    {\it Predicting the behavior of techno-social systems},    {\it Science}  {\bf 325}, 425-428  

\bibitem{stanley}
Stanley H E, Amaral  L A N, Buldyrev S V, Goldberger A L, Havlin S,  Leschhorn H, Maass P, Makse H A, Peng C.-K, Salinger M A, Stanley M H R  and Viswanathan G M  1996    {\it Scaling and universality in animate  and inanimate systems},    {\it Physica A}  {\bf  231}, 20-48 

     \bibitem{[5]YMIoannides}  Ioannides Y  M and  Overman H G  2003        {\it   Zipf's law for cities: an empirical examination},   {\it Regional Science and Urban Economics} {\bf  33} 127-137   
     
\bibitem{newman}
Newman M  E  J  2005    {\it Power laws, Pareto distributions and Zipf's law},   {\it   Contemp. Phys.}  {\bf 46}, 323-351 

\bibitem{g2}
 Gabaix X, Gopikrishnan P,   Plerou V  and  Stanley H  E   2003    {\it A theory of power-law distributions in financial market fluctuations},   {\it Nature} {\bf  423}, 267-270 

\bibitem{z1}
Zipf  G K  1949  {\it Human Behavior and the Principle of Least Effort : An Introduction to Human Ecology} (Cambridge, Mass : Addison Wesley)

\bibitem{Titchmarsh}  Titchmarsh E C 1951 {\it The Theory of the Riemann Zeta Function}  (Oxford University
Press)

\bibitem{s1}
 Sheppard   E  1982    {\it Budget size distributions and spatial economic change } WP-82-31, Working papers of the International Institute for Applied System Analysis, Laxenburg, Austria  
 
       \bibitem{PhA391.12.767ranksizescaling}   Yanguang Chen  2012   {\it   The rank-size scaling law and  entropy-maximizing principle},  {\it   Physica A} {\bf 391},  767-778

\bibitem{gibrat}
Gibrat R  1957    {\it   On economic inequalities},  {\it    International Economic Papers} {\bf  7},  53-70 

 \bibitem{JAS24.97.635worldsportsratingStefani}  Stefani  R T  1997    {\it Survey of the major world sports rating systems},   {\it J. Appl. Stat.} {\bf  24},  635-646  

   \bibitem{interfaces35.06.497NCAA}  
  Cassady C  R, Maillar L M and Salman S  2005      {\it Ranking Sports Teams: A Customizable Quadratic Assignment Approach},   {\it  Interfaces} {\bf 35}, 497-510  

\bibitem{COR33.06.2057}   Churilov L  and Flitman A  2006    {\it Computers \& Operations Research},
{\bf 33}, 2057-2082  

 \bibitem{owgr_20120507_broadie_rendleman}
Broadie M and Rendleman R J 2013  {\it Are the official world golf rankings biased?},
  {\it Journal of Quantitative Analysis in Sports},
 {\bf 9} (2), 127--140. 
 
\bibitem{NJP14.12093038scoreprize}   Deng Weibing,   Li Wei, Cai  Xu,  Bulou A  and Wang Qiuping A 2012
 {\it Universal scaling in sports ranking},   {\it New J. Phys.}  {\bf 14},  093038,
 
 \bibitem{pilavci2011pitch}  Pilavci B 2011 {\it On-pitch success in UEFA Champions League: an empirical analysis of economic, demographic and traditional factors}, Master's  thesis within Economics and Management of Entertainment \& Art Industries, J\"{o}nk\"{o}ping University.
   
 
   \bibitem{JSE10.09.582weightranking} 
  Fainmesser I,   Fershtman Ch and Gandal  N  2009    {\it A Consistent Weighted Ranking Scheme With an Application to NCAA College Football Rankings},   {\it J. Sports Econ.}  {\bf  10}, 582-600  

   \bibitem{NFLranking} Govan A Y and  Meyer C D  2006   {\it  Ranking National Football League Teams using Google's Pagerank},   {\it Center for Research in Scientific Computing, North-Carolina State University, Raleigh, NC} pp. 27695--8205 
   $http://www ncsu edu/crsc/reports/ftp/pdf/crsc-tr06-19 pdf$
   
 \bibitem{DAM108.01.317}  Kern W  and   Paulusma  D   2001    {\it The new FIFA rules are hard: complexity aspects of sports competitions},   {\it Discr.  Appl. Math.} {\bf 108}, 317-323. 
           
           \bibitem{JSE08.07.202explainingrankingsoccercountries}  Macmillan P   and Smith I   2007    {\it Explaining International Soccer Rankings},   {\it J. Sports Econ.} {\bf  8}, 202-213. 


          \bibitem{IJMPCFIFAMARCAGNV} Ausloos M, Cloots R, Gadomski A and Vitanov N K  2014 {\it Ranking structures and Rank-Rank Correlations  of  Countries.  The FIFA and UEFA cases},  {\it  arxiv 1403.5683}

  \bibitem{pi-ratings} Constantinou A C and Fenton N E
2013  {\it Determining the level of ability of football teams by dynamic ratings based on the relative discrepancies in scores between adversaries}, {\it Journal of Quantitative Analysis in Sports},
 {\bf 9}(1),  37--50. 
           
\bibitem{Ranvaudflagerrors} 
  Baldo M V C,  Ranvaud  R D, and   Morya  E  2002     {\it Flag errors in soccer games: the flash-lag effect brought to real life},   {\it  Perception} {\bf 31},   1205-1210.
 
  \bibitem{JAS36.09.soccer}   Skinner G K  and Freeman G H  2009    {\it Soccer matches as experiments: how often does the 'best' team win?},   {\it J. Appl.  Stat.}  {\bf  36},  1087-1095.
  
 \bibitem{0608007} Ben-Naim E,  Vazquez F, and Redner S  2006 {\it Parity and predictability of competitions}, {\it Journal of Quantitative Analysis in Sports},
 {\bf 2} (4), 1--12.
  
  \bibitem{IJSS3.13.102goalsEuro012} Leite  W S S  2013    {\it Euro 2012: analysis and evaluation of goals scored},   {\it  Int. J. Sports Science} {\bf 3}, 102-106 

  \bibitem{EPJB67.09.459Janke} Bittner E,  Nussbaumer A, Janke W,  and Weigel M   2009    {\it Football fever: goal distributions and non-Gaussian statistics},   {\it   Eur.  Phys. J.  B}  {\bf 67},  459-471 
  
    \bibitem{EPhL78.07.58002Janke}  Bittner E,  Nussbaumer A, Janke W,  and Weigel M  2007    {\it Self-affirmation model for football goal distributions},   {\it  Europhys.  Lett.}  {\bf 78},  580024 
  
   \bibitem{12076796Nevo}  Nevo D  and Ritov Y    2013    {\it Around the goal: Examining the effect of the first goal on the second goal in soccer using survival analysis methods},  {\it   Journal of Quantitative Analysis in Sports}  {\bf 9} (2), 165--177. 
  
 \bibitem{EPJB67.09.445Heuer}   Heuer A and  Rubner O 2009   {\it  Fitness, chance, and myths: an objective view on soccer results},  {\it   Eur.  Phys. J.  B}  {\bf 67}  445-458 

    \bibitem{EPJB75.10.327soccermendes}   Ribeiro  H V, Mendes, R S,  Malacarne L C,   Picoli S  Jr, and Santoro P A  2010   {\it   Dynamics of tournaments: the soccer case},   {\it   Eur.  Phys. J.  B}   {\bf 75},  327-334 
   
 \bibitem{JSMSEvolutionWorldCupfinal}  Wallace J L and Norton  K I  2013  {\it   Evolution of World Cup soccer final games between 1966-2010: Game structure, speed and play patterns},   {\it J. Sci. Med.  Sport}  in press 
   
   \bibitem{soc}  Bak P,   Tang C,  and   Wiesenfeld K 1987 
   {\it Self-organized criticality: An explanation of 1/f noise}
    {\it Phys.  Rev.  Lett. },  {\bf 59},
381-384   

   \bibitem{roehner2007driving} Roehner B M  2007 {\it Driving forces in physical, biological and socio-economic phenomena: a network science investigation of social bonds and interactions},
  (Cambridge University Press).
 
   \bibitem{KT73}  Kosterlitz J M and Thouless D  J  1973 {\it Ordering, metastability and phase transitions in two-dimensional systems},  {\it Journal of Physics C: Solid State Physics} {\bf  6}, 1181-1203
 
 \bibitem{Levenberg44} Levenberg K 1944  {\it   A method for the solution of certain problems in least squares},  {\it   
Quart. Appl. Math.} {\bf 2}, 164-168.
  
 \bibitem{Marquardt63}  Marquardt DW 1963 {\it An Algorithm for Least-Squares Estimation of Nonlinear Parameters}
{\it Journal of the Society for Industrial and Applied Mathematics}, {\bf 11}(2), 431-441. 

\bibitem{LMalgorithm}  
 Lourakis M I A  2011 {\it A Brief Description of the Levenberg-Marquardt Algorithm Implemented by levmar},Ê {\it Foundation of Research and Technology}, {\bf 4}, 1--6.
 
\bibitem{ranganathan2004levenberg} Ranganathan A  2004 {\it The Levenberg-Marquardt algorithm}, {\it Tutorial on LM Algorithm} 

\bibitem{AoyamaTakayasu} Aoyama  H,  Souma W, Nagahara Y, Okazaki M P,  Takayasu H, and Takayasu  M 2000
  {\it Pareto's law for income of individuals and debt of bankrupt companies},
  {\it Fractals}, {\bf 8} (3),  293--300.
  
\bibitem{gravetter2013essentials} Gravetter F and Wallnau L  1013 {\it Essentials of statistics for the behavioral sciences}, (Cengage Learning), ch.8
   
        \bibitem{FAIRTHORNE}  Fairthorne R A  1969  {\it  Empirical hyperbolic distributions (Bradford-Zipf-Mandelbrot) for bibliometric description and prediction},
  {\it  Journal of Documentation}, {\bf  25},  319-343. 

     \bibitem{Pwco3}    Rose C,  Murray D, and    Smith D   2002 {\it  Mathematical Statistics with Mathematica}, Springer, New York,   p. 107.
     
     \bibitem{Sofia3a}  Ausloos  M   2013    {\it   A scientometrics law about co-authors and their ranking  The co-author core},  {\it    Scientometrics} {\bf  95}, 895-909  
    
 \bibitem{MABSS} 
Ausloos M 2014 {\it Binary Scientific Star Coauthors Core Size},  {\it Scientometrics},  1-21; $arxiv $ $ 1401.4069$
  
  \bibitem{[HB]}  
 Bougrine  H 2014 {\it  Subfield Effects on the Core of Coauthors},   {\it  Scientometrics}Ê{\bf   98}(2), 1047-1064.
 
    \bibitem{[JM]}  
 Miskiewicz J 2013  {\it Effects of Publications in Proceedings  on the Measure of the Core Size of Coauthors},  {\it Physica A} Ê{\bf  392}  (20), 5119-5131.  
 
  \bibitem{[GR]}    Rotundo G 2014  {\it Black-Scholes-Schršdinger-Zipf-Mandelbrot model
framework for improving a study of the coauthor core score}, {\it Physica A} Ê{\bf  404}, 296-301.


\bibitem{Brakman} Brakman S, Garretsen H, van Marrewijk, C  and van den Berg M  1999 {\it    The  Return of Zipf: Towards a Further Understanding of the Rank-Size Distribution}, {\it   Journal of  Regional Science} {\bf 39}, 182-213 

 \bibitem{dissip}  Nicolis G  and Prigogine I 1977  {\it    Self-Organization in Nonequilibrium
Systems: From Dissipative Structures to Order Through Fluctuations}    (Wiley, New York)

 \bibitem{BDMaz47}    Bedeaux D and   Mazur   P
           1981 {\it     Stability of Inhomogeneous Stationary States for the hotspot Model of a Superconducting Microbridge}, {\it   Physica A} {\bf  105}, 1-30  
           
\bibitem{BDMaz46}  Mazur P and    Bedeaux  D  1981
                {\it   An Electro-Thermal Instability in a Conducting Wire: Homogeneous and Inhomogeneous Stationary States for an Exactly Solvable Model},  {\it                   J.  Stat.  Phys. } {\bf 24}, 215-233  
                
\bibitem{PhB108hotspot}    Ausloos M 1981 {\it Continuously Forded Ballast Resistor Model for Superconducting Hot Spots},  {\it   Physica B} {\bf  108}, 969-970      
  
  \bibitem{[9]}  Klimenko  A Y and   Pope S B   2012 {\it Propagation speed of combustion and invasion waves
in stochastic simulations with competitive mixing}, {\it  Combustion Theory and Modelling}
{\bf 164} 679-714   

  \bibitem{[10]}  Klimenko A Y 2012  {\it Mixing, entropy and competition},    {\it Physica Scripta }Ê {\bf 85},  068201 

 \bibitem{tellus11.59.267.kuettnercloudbands} Kuettner J 1959 {\it The band structure of the atmosphere}, {\it Tellus} {\bf 11},  267--294  

\bibitem{romanovsky1940application} Romanovsky  V 1940 {\it Une application des tourbillons convectifs  Formation des sols polygonaux}, {\it J.  Phys.  Radium}, {\bf 1}, 346--349 

\bibitem{romanovsky1941application} Romanovsky V 1941   {\it Application du crit{\'e}rium de Lord Rayleigh {\`a} la formation des tourbillons convectifs dans les sols polygonaux du Spitzberg},  {\it C. R. Acad.  Sci.  Paris} 
 {\bf 211}, 877--878 
 
   \bibitem{[24]}  Carath\' eodory C   1976 {\it  Investigations into the foundations of thermodynamics}, {\it The Second Law of Thermodynamics}  {\bf  5}, 229--256 

\bibitem{[25]} Lieb  E  H and Yngvason J 
 1999  {\it The physics and mathematics of the second law of thermodynamics}, {\it   Phys.  Rep.} {\bf  310}, 1-96  
 
  \bibitem {caram2010dynamic}  Caram  L,  Caiafa C, Proto A,  and Ausloos M 2010 {\it  Dynamic peer-to-peer competition}, {\it  Physica A} {\bf  389}(13), 2628-2636  

  \bibitem{Merton}  Merton R K 1968    {\it The Matthew Effet}, {\it Science}  {\bf 159},  56-63.

   \end{thebibliography}
    \end{document}